\documentclass[aps,prd,reprint,nofootinbib,longbibliography]{revtex4-1}
\usepackage{blindtext}
\usepackage{mathtools}
\usepackage{xcolor}
\usepackage{adjustbox}
\usepackage{graphicx}

\usepackage{soul}

\usepackage[utf8]{inputenc}
\usepackage[english]{babel}
\usepackage{hyperref}
\hypersetup{
    colorlinks=true,
    linkcolor=blue,
    citecolor=red,
    filecolor=magenta,      
    urlcolor=magenta,
}

\def\sfrac#1#2{\ensuremath{\textstyle \frac{#1}{#2}}}

\begin{document}
\title{Can primordial black holes as all dark matter explain fast radio bursts?}
\author{Kimmo Kainulainen$^{1,2*}$, Sami Nurmi$^{1,2**}$, Enrico D. Schiappacasse$^{1,2\dagger}$, Tsutomu T. Yanagida$^{3\ddagger}$}

\affiliation{
$^1$ Department of Physics, University of Jyv$\ddot{a}$skyl$\ddot{a}$, P.O.Box 35 (YFL), FIN-40014 Jyv$\ddot{a}$skyl$\ddot{a}$, Finland\\
$^2$Helsinki Institute of Physics, University of Helsinki, P.O. Box 64, FIN-00014 Helsinki, Finland\\ 
$^3$Tsung-Dao Lee Institute and School of Physics and Astronomy, Shanghai Jiao Tong University, 
 200240 Shanghai, China }

\begin{abstract}
Primordial black holes (PBHs) are one of the most interesting nonparticle dark matter (DM) candidates. They may explain all the DM content in the Universe in the mass regime from about $10^{-14}M_{\odot}$ to $10^{-11}M_{\odot}$. We study PBHs as the source of fast radio bursts (FRBs) via magnetic reconnection in the event of collisions between them and neutron stars (NSs) in galaxies. We  investigate  the energy loss of PBHs during PBH-NS encounters to model  their capture by NSs. To an order-of-magnitude estimation, we conclude that the parameter space of PBHs being all DM is accidentally consistent with that to produce FRBs with a rate which is the order of the observed FRB rate.
\end{abstract}

\maketitle

\section{Introduction}

One of the outstanding challenges facing the physics community is the unknown identity of  dark matter (DM).
Even though its existence has been confirmed by astrophysical and cosmological observations, we do not know much about its properties apart from the fact that it interacts at least gravitationally~\cite{Bertone:2016nfn, Freese:2017idy}. Among the most favored particle DM candidates,   
the weakly interacting massive particles~\cite{Roszkowski:2017nbc} and the axion~\cite{PRESKILL1983127, ABBOTT1983133, DINE1983137, Kim:2008hd}
remain elusive after three decades of collider and direct DM searches.
Another interesting class of DM is  
astronomical objects such as primordial black holes~\cite{Hawking:1971ei} (PBHs), formed 
in the very early Universe. The study 
of PBHs dates back to the 1960s, and it was soon realized~\cite{1975Natur.253..251C} that they constitute an excellent DM candidate  with a rich phenomenology 
~\cite{Kawasaki:1997ju, Khlopov_2010, Carr:2016drx,Eroshenko:2016yve, Hertzberg:2020hsz, Hertzberg:2020kpm, Nurmi:2021xds, Natwariya:2021xki, Ashoorioon:2020hln}.
For sufficiently large masses PBHs are stable and behave like cold DM on cosmological scales.
The possibility that
PBHs constitute all of the DM has already been ruled out by dynamical and relic observations,
except for the mass range
$10^{-14} M_{\odot}-10^{-11}M_{\odot}$,  where current bounds are subject to uncertainties~\cite{PhysRevLett.80.1138, Takahashi:2003ix, Carr:2020gox}. Indeed, a successful inflation model was proposed by one of us in Refs.~\cite{Inomata:2017okj, Kawasaki:2016pql} to explain all DM with a sharp peak in the PBH mass spectrum around $10^{-13}M_{\odot}$.
This PBH all-DM hypothesis has  
the shocking consequence that we do not need to go beyond the Standard Model of particle physics to explain the DM content of the Universe. 
In this article, we adopt this hypothesis and study if such PBHs making up the DM component could also be the the source of fast radio bursts (FRBs), reproducing the correct observed rate. 

FRBs are bright short duration pulses of emission in the radio frequency spectrum, whose 
origin is at present unknown. The first detection of a FRB event was reported by Lorimer and collaborators~\cite{2007Sci...318..777L} 
based on the
archival survey data from the Parkes radio telescope. 
Since then, 
the physics community has made great efforts 
to confirm and explain the 
phenomenon. There is no a standard criteria  
to define a FRB event, but this is a matter of ongoing research~\cite{2018MNRAS.481.2612F}. In practice, a signal is identified as a FRB if it satisfies a set of relaxed criteria including pulse duration, brightness, and a dispersion measure. 

While the typical FRB duration is the order of milliseconds or less, their radiated power, up to  $\sim 10^{43}\,\text{erg/s}$ (assuming an isotropic emission) is comparable to that of an entire typical radio galaxy. These brief and energetic events could be  linked to neutron stars (NSs), which 
feature 
large gravitational, magnetostatic and electromagnetic energy densities with short characteristic timescales: $\mathcal{O}(2G_N M_{N})\sim 10^{-2}\,\text{ms}$, using $M_{N}\sim M_{\odot}$ as a typical NS mass scale. Indeed, the extremely high brightness temperature associated with FRB events (temperature of the Wien's law 
fitting the
observed flux density)
could be an indication of the presence of plasma and/or magnetic fields~\cite{PhysRevD.89.103009}.
Many models involving NSs 
have already been proposed to explain FRBs. These range from catastrophic events, such as colliding or merging NSs~\cite{Totani:2013lia, Lipunov:2013axa, Yamasaki:2017hdr}, NS collapse~\cite{2014ApJ...780L..21Z, DasGupta:2017uac}, \textcolor{black}{NS and dense axion star collisions~\cite{Buckley:2020fmh}}, NS and asteroid collisions, NS 
quakes, and lightning
in the NS magnetosphere to interaction of a pulsar with its environment~\cite{2017ApJ...838L...7D, Waxman:2017zme} or the model of soft gamma repeaters~\cite{Metzger_2017, Wang_2017} (for a review, 
see Ref.~\cite{Katz:2018xiu}). 

In Ref.~\cite{Abramowicz_2018}, 
encounters between NSs and PBHs were proposed
as the fast radio burst engine. The NS would capture a PBH due to its dense neutron medium, 
and as the PBH eventually swallows its host, 
a FRB is formed due to the release of the magnetic field energy during the final collapse of the NS to a black hole (BH)~\cite{Fuller:2017uyd}. This mechanism is consistent with the typical short duration and high luminosity characteristics of FRBs.  However, results from~\cite{Abramowicz_2018} suggest 
that the PBH all-DM scenario with PBH masses in the range of $10^{-14} M_{\odot}-10^{-11}M_{\odot}$ is in tension with observations, producing FRBs with a rate that is 
several orders of magnitude too large  
(assuming the observed signal is uniformly distributed among $10^{11}$ galaxies). In this article, we argue  
that these results are based on an overestimation of the actual PBH capture rate by NSs and that with more appropriate rates 
the PBH all-DM hypothesis, quite remarkably, predicts 
a FRB rate of the correct magnitude. 

\section{FRB ocurrence rate: numerical simulation}

If all DM in galactic halos is in the form of PBHs, they will unavoidably collide with NSs. Under the right conditions, NSs may capture PBHs, which eventually leads to $\text{NS} \rightarrow \text{BH}$ conversion and the emission of the original NS magnetic field energy. Such released energy would power the FRB events. 

\begin{figure}[t!]
\centering
\includegraphics[width=7.5 cm]{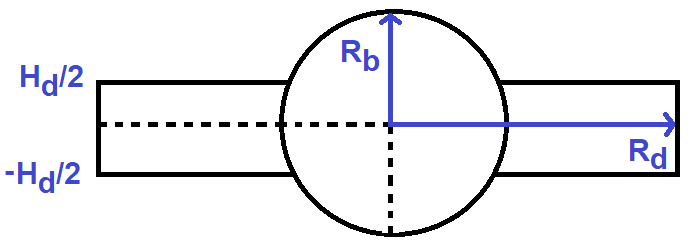}
\caption{Toy model for a typical galaxy composed of a spherical bulge with radius $R_b = 5.0\, \text{kpc}$ and disc plane with height $H_d = 1.0\, \text{kpc}$ and radius $R_d=30\,\text{ kpc}$.} 
\label{Plottoy}
\end{figure}

For the sake of comparison, we follow the simple order-of-magnitude estimation performed in Ref.~\cite{Abramowicz_2018} to simulate the galactic halo interaction with the bulge and the disk of a typical spiral galaxy.  Thus, we define a toy model galaxy parametrized by the galactic disk thickness, $H_d$, galactic disk radius, $R_d$, and spherical galactic bulge radius, $R_b$, as seen in Fig.~\ref{Plottoy}.
The density $\rho(r)$ within a galactic radius $r$ is taken to be 
\begin{equation}
\rho(r) = 
\begin{cases}
\frac{v_d^2}{4\pi G_N r^2} \,\,\text{for}\,\,  r > R_b\,, \\
\,\hspace{3.7 cm} \\
\frac{3v_d^2}{4\pi G_N R_b^2}\,\,\text{for}\,\, r \leq R_b\,,
\end{cases}
\label{relaxed}
\end{equation} 
such that it obeys a simplified rotational curve where $v(r > R_b) = v_d$ and $v(r \leq R_b) =  v_d r/R_b$,
with $v(r)$ being the radial circular velocity and $v_d = 220\,\text{km/s}$. The sum of the masses of the disk and bulge is obtained as
\begin{align}
\hspace{-0.6cm}M_{\text{total}}& =  \int_0^{R_b} \hspace{-0.2 cm}4\pi r^2 \rho(r) dr +  \int_{-\frac{H_d}{2}}^{\frac{H_d}{2}} \int_{R_b}^{R_d}\hspace{-0.2 cm} 2\pi R \rho(r) dR dz\,,\\
& \approx \frac{v_d^2 R_b}{G_N} + \frac{H_d v_d^2}{2 G_N}\text{ln}\left( \frac{R_d}{R_b} \right)\sim 7\times 10^{10}\,M_{\odot}\,,
\end{align}
where $r=\sqrt{R^2+z^2}$ in cylindrical coordinates $(R,\phi,z)$.

Our scenario has several interacting astrophysical objects 
in the galaxy. 
Apart from light PBHs and NSs with a typical mass of $M_N = 1.5\,M_{\odot}$, we  eventually have light black holes with a mass $M_{0}= M_{N}$ produced by the $\text{NS} \rightarrow \text{BH}$ 
\begin{table*}[!]
\centering
\begin{tabular}{lllll}
(Index) Type & Mass & Number density & Radius \\
\hline
(N) NS & $M_{N}=1.5 M_{\odot}$ & \hspace{0.8 cm} $n_{N}$ & $R_N=10\,\text{km}$ \\
(P) PBH & $M_{P} = \overline{M}_{\text{PBH}}$ & \hspace{0.8 cm} $n_{P}$ & $R_{P} = R_s(M_{P})$ \\
(0) Light BH & $M_0=1.5 M_{\odot}$ & \hspace{0.8 cm} $n_{0}$ & $R_0=R_s(M_0)$  \\
(1) Stellar PBH & $M_1 =10 M_{\odot}$ & \hspace{0.8 cm} $n_{1}$ & $R_1=R_s(M_1)$ 
\end{tabular}
\caption{Different compact objects of interest in the galaxy: masses ($M_i$), number densities ($n_i$), and radii ($R_i$). The Schwarzschild radius of the i\textit{th} species is given by $R_s(M_i) = 2G_N M_i$.}
\label{tab:my-table}
\end{table*}
conversion, and we add a population of stellar black holes (BHs) as a result of standard stellar evolution.
We summarize the main parameters 
in our simulation in Table~\ref{tab:my-table}. Also the labeling used below for the various species is explained in the table. 
If the DM is composed of a nonmonochromatic PBH mass spectrum, as in the inflation model  proposed in Ref.~\cite{Inomata:2017okj},  we use $M_P = \overline{M}_{\text{PBH}}$, where $\overline{M}_{\text{PBH}}$  is the mean PBH mass. As long as $M_{N}\gg \overline{M}_{\text{PBH}}$, which is the case  here,  
this is practically equivalent to using the full extended mass spectrum. 

We use  the following initial number densities in terms of the galactic radius:
\begin{align}
n_{P}(t=0,r) & = \frac{\rho(r)}{{\color{black}M_{\text{P}}}}\,,
\label{np1}\\
n_0(t=0, r)&=n_1(t=0, r)=n_N(t=0, r) = 0\,,
\end{align}
where $n_{P}$ is the number density of PBHs. In addition, $n_0$, $n_1$, and $n_N$ are the number density of light BHs after NS conversion, stellar BHs, and NSs, respectively. 

The velocity averaged collision rate between species $i$ and $j$ contains an important enhancement
due to gravitational focusing which deflects the trajectories of the colliding objects (see, for example, Sec.~4.1 in Ref.~\cite{Hertzberg:2020dbk}):
\textcolor{black}{
\begin{align}
&C^{ij}(r)n_i(r)n_j(r) = \langle \sigma_{\text{eff}}(v_{\text{rel}})v_{\text{rel}}(r) \rangle n_i(r)n_j(r)\,,\label{Cij}\\
&=  \pi (R_i + R_j)^2 n_i(r)n_j(r) \times\nonumber\\
&\int d^3v_{\text{rel}} \left( 1 + \frac{2G_N (M_i + M_j)}{(R_i + R_j)v_{\text{rel}}^2} \right)v_{\text{rel}}P(v_{\text{rel}})\,,\label{Cijdetail}
\end{align}}
%
 \hspace{-0.2 cm} where $P(v_{\text{rel}})dv_{\text{rel}}$ denotes the probability for the relative velocity be in the interval $v_{\text{rel}} ...v_{\text{rel}} + dv_{\text{rel}}$.  We assume the   Maxwell-Boltzmann distribution for all species,  
\textcolor{black}{
\begin{equation}
 P(v_{\text{rel}})dv_{\text{rel}} = \left(\frac{3}{2\pi\sigma_{\text{rel}}^2}\right)^{3/2}  \text{exp}\left(-\frac{3 v_{\text{rel}}^2}{2\sigma_{\text{rel}}^2}\right)\,, 
\end{equation}
where we assume $\sigma_{\text{rel}}$ is the order of $v(r)$, the rotational velocity corresponding to Eq. (\ref{relaxed}).
For example, for $i=N$ and $j=P$, we have $M_i \gg M_j$ and $R_i \gg R_j$ with the gravitational enhancement dominating the effective cross section.  Calculating the integral in Eq.~(\ref{Cijdetail}), we have
%
%
\begin{align}
C^{NP_1}&(r)n_N(r)n_{P_1}(r) \simeq \textcolor{black}{1.6\times 10^{-12}}\,\text{yr}^{-1}\times \nonumber \nonumber\\
&\times \Big(\frac{M_N}{\textcolor{black}{1.5\,M_{\odot}}} \Big)\Big(\frac{R_N}{10\,\text{km}} \Big) 
\Big(\frac{{\color{black} 300\; {\rm km}/{\rm s}}}{\sigma_{\text{rel}}} \Big) 
\nonumber\\
&\times 
\Big(\frac{\textcolor{black}{10^{-12}M_{\odot}}}{M_{P}} \Big)
\Big(\frac{\rho_{P}(r)}{\text{GeV cm}^{-3}} \Big) n_N(r).
\label{coll1}
\end{align}
Equation (\ref{coll1}) corresponds to the number of NS-PBH collisions per year times the NS number density at a galactic radius $r$. However, in contrast to what was assumed in Ref.~\cite{Abramowicz_2018}, not all NS-PBH collisions lead to a PBH capture by the NS. The initial NS-PBH impact parameter needs to be small enough so that the PBH energy loss during the interaction is large enough to overcome the initial PBH energy, as discussed in detail in  Ref.~\cite{Genolini:2020ejw}. Suppose that there is a background of PBHs with a relative velocity dispersion $\sigma_{\text{rel}}$ at infinity with respect to an isolated NS. The capture condition during a PBH encounter reads 
\begin{equation}
|\Delta E| > E_i = \sfrac{1}{2}m \sigma_{\text{rel}}^2\,,\label{econd} 
\end{equation}
where the total energy loss
$|\Delta E| = |\Delta E^{in}_{\text{GW}}| + |\Delta E_{\text{GW}}^{out}| + |\Delta E_{\text{surf}}| + |\Delta E_{\text{dyn}}|$ 
consists of the loss via dynamical friction $|\Delta E_{\text{dyn}}|$,  the energy radiated in gravitational waves $|\Delta E_{\text{GW}}|$ during the PBH motion inside ($in$) and outside ($out$) the NS and the energy dissipation by the PBH due to the production of surface waves during the NS crossing $|\Delta E_{\text{surf}}|$. For typical NS parameters, the ratio between the capture and collision rates for the interaction between an isolated NS and a background of PBHs moving with a dispersion velocity $\sigma_{\text{rel}}$ in the NS frame is given by~\cite{Genolini:2020ejw} 
\begin{align}
F_{\text{cap/coll}} &\approx \textcolor{black}{ 10^{-5}} \frac{\text{captures}}{ \text{collisions}}\times\,\nonumber\\
&\left( \frac{{\color{black} 300\; {\rm km}/{\rm s}}}{\sigma_{\text{rel}}} \right)^2 \left( \frac{\textcolor{black}{M_{P}}}{\textcolor{black}{10^{-12}\,M_{\odot}}} \right) \mathcal{C}[X]\,,\label{fap/coll}
\end{align}
where $X \equiv \textcolor{black}{2\times 10^{-4}(10^{-12}M_{\odot}}/\textcolor{black}{M_{P}})^{-1}({\color{black} 300\; {\rm km}/{\rm s}}/\sigma_{\text{rel}})^2$. The function $C[X] = 1$ for $X \lesssim 10$ and it declines as $C[X]\sim X^{\alpha}$ for larger $X$, with $\alpha = -1$ for $10 \lesssim X \lesssim 1000$ and $\alpha = -5/7$ for $X \gtrsim 1000$ (see Fig.~3 in Ref.~\cite{Genolini:2020ejw}).}
\textcolor{black}{The collision rate expressed in Eq.~(\ref{coll1}) is transformed to capture rate by using Eq.~(\ref{fap/coll}) as
\begin{align}
\widetilde{C}^{NP}&(r) n_N(r) n_{P}(r) =\,\nonumber\\
&F_{\text{cap/coll}}(v(r),\textcolor{black}{M_{P}}) C^{NP}(r) n_N(r) n_{P}(r)\,,\label{13}
\end{align}
where we have used the galactic circular velocity 
to estimate
the relative PBH/NS velocity, setting $\sigma_{\text{rel}} = v(r)$}. 

 Neutron stars are born in the core-collapse events of massive stars corresponding to supernovae of types Ib, Ic, and II and, less frequently, of type Ia (accretion-induced collapse of white dwarfs that have reached the Chandrasekhar’s limit)~\cite{2010A&A...510A..23S}. The present day types II + Ib supernovae rate derived from the Galactic $^{26}A1$ mass is  $3.4 \pm 2.8$ per century~\cite{Timmes:1997ua}. In Milky Way-like galaxies, the type Ia supernovae rate, mainly dominated by the double degenerate scenario,  is $\sim 10^{-3}\text{yr}^{-1}$~\cite{2009ApJ...699.2026R}.
 We use such present-day supernovae rates to roughly estimate the total number of NSs generated per a typical spiral galaxy. We take a constant total creation rate for NSs and stellar black holes in the range $K^N=K^1=[0.007-0.063]\, \text{yr}^{-1}$ and weight both of them via the local density as $K^i(r) = (K^i/M_{\text{total}})\rho(r)$, where $i=\{N,1\}$.
 
The time evolution of the number densities of different astrophysical species
follows from a
system of coupled differential equations as 
%
\begin{align}
\dot{n}_N   = -& \;\textcolor{black}{\widetilde{C}^{NP}n_Nn_{P}} -\sfrac{1}{2}C^{NN}n_{N}n_{N}- C^{N0}n_N n_0\,
\nonumber \\ - & \;C^{N1}n_N n_1+ K^N\,,
\label{NS} \\
\dot{n}_{P} = -& \;\textcolor{black}{\widetilde{C}^{NP}n_N n_{P}} -\sfrac{1}{2}C^{PP}n_{P}n_{P} - C^{P0}n_{P}n_0
\nonumber \\  -& \;C^{P1}n_{P}n_1 \,,
\label{p1} \\
\dot{n}_{0} = \phantom{-} &\; \textcolor{black}{\widetilde{C}^{NP}n_N n_{P}}-C^{01}n_0 n_{1}-\sfrac{1}{2} C^{00}n_0 n_{0}
\label{n0} \\
\dot{n}_{1} = -&\sfrac{1}{2}C^{11}n_1n_1 + K^1\,.
\label{n1}
\end{align}
In Eqs.~(\ref{NS})-(\ref{n1}), a prefactor 1/2 is introduced to avoid double counting in
collisions between objects of the same species: PBHs, NSs, light, and stellar BHs.
\begin{figure}[]
\centering
\includegraphics[width=7.5 cm]{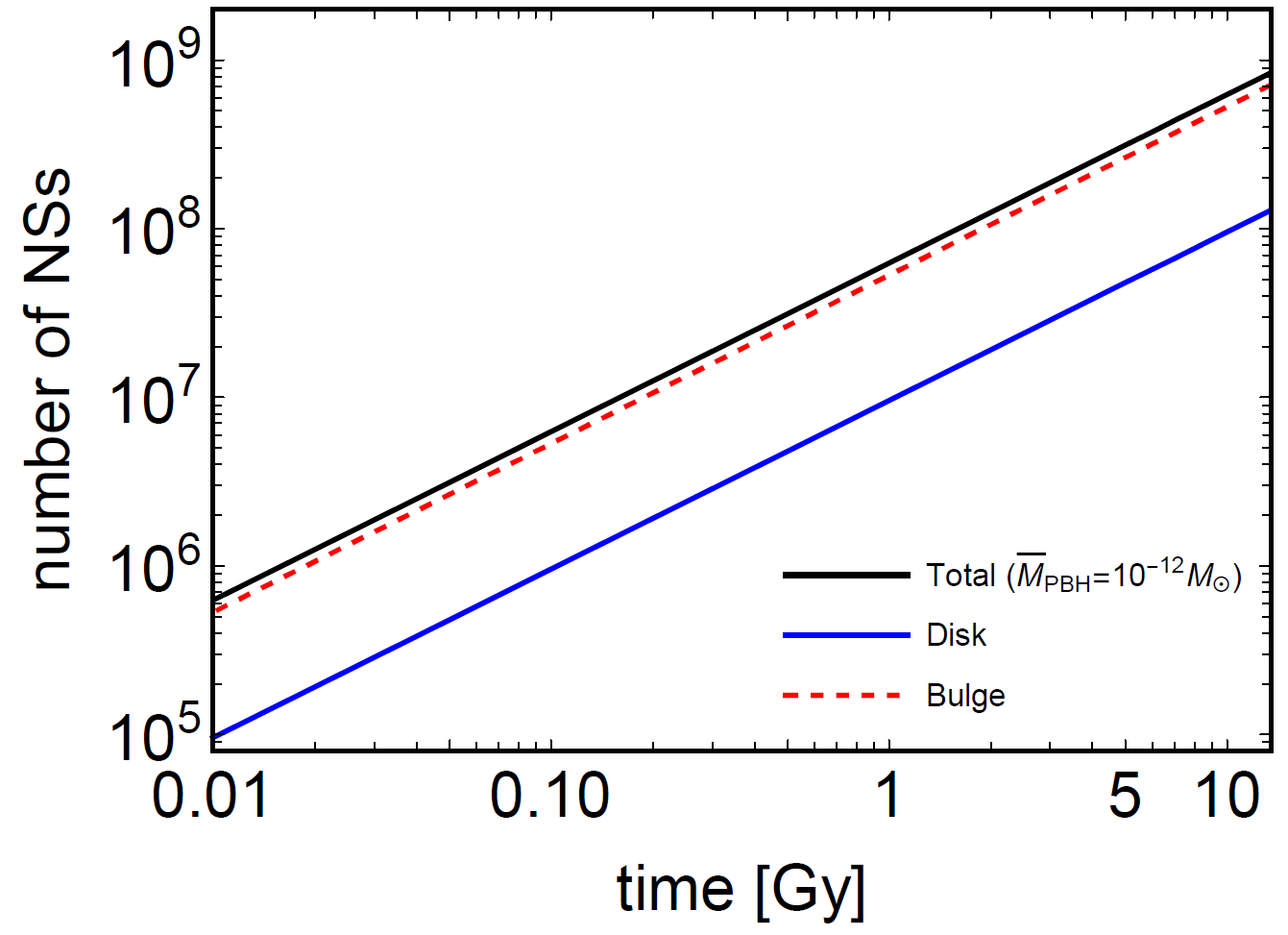}\;
\caption{The total number of NSs (black line) consisting from contributions from the galactic bulge (red dashed line) and disk (blue line) in a typical spiral galaxy.}
\label{Number}
\end{figure}

Figure~\ref{Number} shows the number of NSs as a function of the age of a galaxy for an example with  $\textcolor{black}{M_{P}}=10^{-12}M_{\odot}$ and $K^N=K^1=0.063\,\text{yr}^{-1}$.
The total number of NSs and the contributions from the bulge and disk are shown by 
the black, dashed red, and blue lines, respectively.
The predicted present number of galactic NSs agrees with the total number estimated in the literature~\cite{2010A&A...510A..23S}.

\begin{figure}[t!]
\centering
\includegraphics[width=8.4 cm]{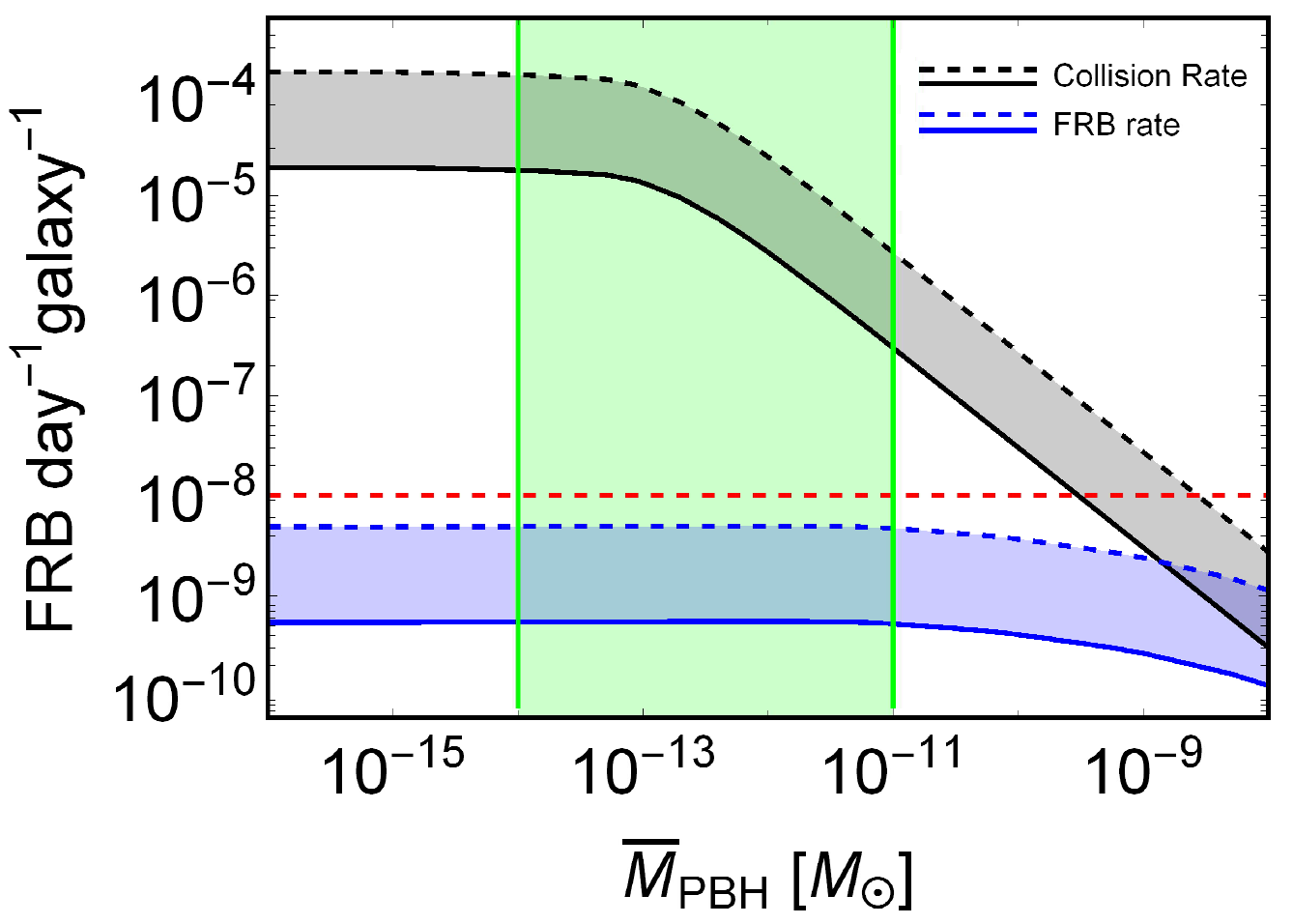}
\caption{Predicted FRB rate in a typical spiral galaxy for $K^N=K^1=0.063\,\text{yr}^{-1}$ (dashed blue line) and for $0.007\,\text{yr}^{-1}$ (solid blue line).  Also shown are the total rates of NS-PBH collisions (dashed black line and solid black lines) and the
observed FRB rate $\sim 10^{-8} \text{galaxy}^{-1} \text{day}^{-1}$ (dashed red line)~\cite{2016MNRAS.460L..30C, Petroff:2019tty}.
 The green shaded region indicates the parameter space where PBHs may still make up all DM.} 
\label{FRB}
\end{figure}  

We calculate the current FRB rate  per galaxy by evaluating $\int \widetilde{C}^{NP_1}(r)n_N(r,t_0)n_{P_1}(r,t_0) d^3r$ over the galactic bulge and disk, using $t_0=13.5\, \text{Gyr}$ that equals the typical galactic age. Assuming that FRB are detected from about $10^{11}$ galaxies in the observable Universe (most FRBs are located at redshifts $\lesssim 1$), the estimated rate of FRBs from observations is $\sim10^{-8}\,\text{day}^{-1}\text{galaxy}^{-1}$
~\cite{2016MNRAS.460L..30C, Petroff:2019tty}. 

Figure \ref{FRB} shows the current number of FRBs  produced in our setup per day and per galaxy in terms of the PBH mass, $\textcolor{black}{M}_{P}$, for $K^N=K^1$ equal to $0.063\,\text{yr}^{-1}$ (dashed blue line) and $0.007\,\text{yr}^{-1}$ (solid blue line). In the parameter space of interest, where PBHs may still constitute all DM, $10^{-14}M_{\odot}\lesssim {\color{black}M_{\text{P}}}\lesssim 10^{-11}M_{\odot}$ (green shaded region), the generated FRB rate is the order of the observed FRB rate (dashed red line). The near constancy of the FRB rate can be understood from Eq.~(\ref{13}). While the capture efficiency increases with the PBH mass,
$F_{\text{cap}/\text{coll}}\propto {\color{black}M_{\text{P}}}$, 
when $\mathcal{C}[X]$ remains a constant, 
this is compensated by the decreasing number of PBHs, $n_{P}\propto {\color{black}M_{\text{P}}^{-1}}$. 
The bending of the rate down at  
${\color{black}{M}_{\text{P}}} \gtrsim 10^{-11} M_{\odot}$ 
comes from the suppression of $\mathcal{C}[X]$ at small galactic radii, where $v(r)\ll v_{\rm d}$ and subsequently $X$ can be very large.

For comparison, we also show in Fig.~\ref{FRB} (the dashed and solid black lines) the FRB rate obtained assuming that every NS/PBH collision leads to a PBH capture, i.e.,~setting $F_{\text{cap/coll}} = 1$,
which was the assumption made in Ref.~\cite{Abramowicz_2018}.
However, as seen in Fig.~\ref{FRB}, this vastly overestimates the actual signal for all masses $M_{\text{P}}\lesssim 10^{-9}M_{\odot}$, and, in particular, for $M_{\text P}\sim10^{-14}M_{\odot}-10^{-11}M_{\odot}$,
it gives a rate that is
several orders of magnitude
above the observed value $\sim 10^{-8}\,\text{day}^{-1}\text{galaxy}^{-1}$.
In contrast, our results show, quite remarkably, that the FRB rate corresponding to the PBHs all-DM scenario
agrees with the observed rate within order of magnitude precision.

\section{Discussion and Outlook}

We have shown that the parameter space where all DM can be PBHs is accidentally consistent with that to produce FRBs with a rate which agrees with observations to the right order of magnitude. Here, we discuss the main uncertainties associated with our estimation and provide some outlook for future research.

The capture rate of PBH by NSs based on Eq.~(\ref{fap/coll}) is calculated  assuming that the NS is a homogeneous sphere of matter. However, NSs have  inhomogeneoeus density profiles with different levels of stiffness depending on the equation of state~\cite{Potekhin:2013qqa}. 
Dynamical friction is expected to be 
enhanced in inhomogeneous setups compared to the homogeneous cases~\cite{Popolo2003}. In our case, enhancement of the dynamical friction would increase both the PBH capture rate by NSs and the resulting FRB rate. 

The FRB rate depends linearly on the NS number. 
We use a simplistic estimate of the NS number 
based on the present-day supernovae core collapse in Milky Way-like galaxies. Uncertainties in the NS creation rate per galaxy will translate into uncertainties in the final FRB rate. 
It has been argued 
that binary black holes formed from Population III (Pop III) stars could explain the mass-weighted merger rate of the LIGO O1 events~\cite{Inayoshi:2017mrs}.
If this is the case, we expect to have in addition NS remnants from Pop III stars. It would be interesting to study how much this would enhance the FRB rate in our setup.  

The predicted FRB rate mainly depends on the PBH/NS interactions in the bulge as the density of NSs is highest  there (see Fig.~\ref{Number}) and the circular velocity decreases towards the galactic center.
The capture efficiency in Eq.~(\ref{fap/coll}) is very sensitive to the velocity of NS/PBH interactions being enhanced for low speed encounters.  In our simplified analysis, we have assumed that DM towards to the galactic center follows the density profile of the galactic bulge with a relative velocity dispersion estimated by the corresponding circular velocity. 
The production of FRBs in the galaxy would increase in models with a larger DM density and/or smaller relative velocity dispersion in the inner galactic region.

Several processes involving NSs have been studied in the literature as possible sources of FRBs. We note that as the NS/PBH collisions studied here convert only a small fraction of NSs into BHs, eventual FRBs from other NS processes should be largely unaffected and could
add on top of the FRB rate studied here.   

Multibody effects could be relevant 
in the dense environment within the galactic disk and bulge. In general, such effects may either
boost or suppress the capture rate~\cite{Genolini:2020ejw}. 
To an order-of-magnitude estimate,  such effects could cause deviations from our analysis for PBH mean masses in the range ~\cite{Montero-Camacho:2019jte}:
\begin{equation}
\label{eq:multibody}
    \textcolor{black}{M_{P}} \lesssim \frac{M_N}{\text{ln}\Lambda} \left( \frac{t_{\text{dyn}}}{t_{\text{tidal}}} \right)^{4/7}\,,
\end{equation}
 where $t_{\text{tidal}} = (G_N M_{\text{halo}}/R_{\text{halo}}^3)^{-1/2}$ is the square root of the inverse tidal tensor, $t_{\text{dyn}}= R_{N}/(G_N M_N/R_N)^{1/2}$ is a timescale associated to PBH energy loss by dynamical friction, and $\text{ln}\Lambda =\text{ln}(b_{\text{max}}/b_{\text{min}})$ is the logarithmic ratio of the maximum and minimum impact parameters, typically in our setup $\text{ln}\Lambda \sim 30$. 
For typical values used in this paper, Eq.~(\ref{eq:multibody}) gives $\textcolor{black}{M_{P}} \lesssim 2\times 10^{-13}\,M_{\odot}$ for the disk in agreement with Ref.~\cite{Montero-Camacho:2019jte} and $\textcolor{black}{M_{P}} \lesssim 7\times 10^{-13}\,M_{\odot}$ for the bulge. This estimate indicates that the lightest extreme of the parameter space of our interest could be affected by multibody effects. 

\textcolor{black}{We will continue this work in an incoming project~\cite{Kimmo:2021} with 
improvements in modeling, such as time dependent NS 
creation rate, state-of-the art galactic model, and PBH velocities distribution, among others.} 


After the PBH is settled within the NS, it 
will accrete the star and convert it to a light black hole. For most of the parameter space, 
the accretion process occurs in the Bondi regime,%
\footnote{Only warmer and rapidly rotating NSs break such regime, but such conditions would be statistically disfavored by the data~\cite{Lorimer:2006qs, Kouvaris:2013kra} 
. }  
which assumes a spherically symmetric accretion and zero vorticity of the matter which is collapsing into the black hole.  The characteristic time $t_B$ for the accretion is given by
\begin{equation}
t_{B}\sim 2.5\,\text{yr}\left( \frac{c_s}{0.6} \right)^3 \Big( \frac{10^{15}\,\text{gr/cm}^{3}}{\rho_c} \Big) \Big(\frac{10^{-12}\,M_{\odot}}{M_{\text{P}}} \Big)\,,
\end{equation}

\hspace{-0.35 cm}\textcolor{black}{where we have used $c_s = 3/5$ as a characteristic value for the sound speed of the matter in the core of the NS.}  
We see that for PBH masses 
$\lesssim 2\times 10^{-13} M_{\odot}$ the Bondi time is  $\gtrsim 10$ years.  The PBHs with lighter masses could be the source of repeating FRBs. A long-lasting accretion process may explain the irregular bursts of the repeating FRB 121102~\cite{Houben:2019xla} through intermittent reconnection of the magnetic field line bundles. Indeed, it has been argued in the literature that all FRBs could actually be repeating events, and the reason for nonobservation of multiple bursts is just due to a low burst rate. A sizeable number of repeating FRBs may be realized for a nonmonochromatic PBH mass spectrum. Currently, some FRBs have little or no follow-up so that more data are needed to settle down the controversy.  

\vspace{0.25 cm}

\section{Acknowledgments}
E.D.S. thanks Pasquale Serpico for useful discussions about the
capture rate of PBHs by NSs.  T. T. Y. thanks Fabo Feng, Yosuke Mizuno, and Weikang Lin at TDLI for valuable discussions.
This work was supported by the Academy of Finland Grant No. 318319.
 T.T.Y. is supported in part by the China Grant for Talent Scientiﬁc
Start-Up Project and the JSPS Grant-in-Aid for Scientiﬁc Research Grants No.
16H02176, No. 17H02878, and No. 19H05810 and by World Premier
International Research Center Initiative (WPI Initiative), MEXT, Japan.\\

\hspace{-0.32 cm}$^*$\href{mailto:kimmo.kainulainen@jyu.fi}{kimmo.kainulainen@jyu.fi}\\
$^{**}$\href{mailto:sami.t.nurmi@jyu.fi}{sami.t.nurmi@jyu.fi}\\
$^\dagger$\href{mailto:edschiap@uc.cl}{edschiap@uc.cl}\\
$\ddagger$\href{mailto:tsutomu.tyanagida@sjtu.edu.cn}{tsutomu.tyanagida@sjtu.edu.cn}

\bibliography{FRBPBH} 

\end{document}